\documentclass[
    final            
  ]
  {aipproc}

\layoutstyle{6x9}
\usepackage{axodraw}
\usepackage{epsfig}
\usepackage{psfrag}
\begin{document}
\begin{flushright}
SHEP 0602
\end{flushright}
\title{Lattice Flavourdynamics~\footnote{Plenary lecture presented at the
2005 Particles and Nuclei International Conference (PANIC05),
Santa Fe, New Mexico, USA, Oct. 24\,--\,28th 2005.}}



\author{Chris T Sachrajda}{
  address={School of Physics and Astronomy, University of Southampton,
  Southampton SO17 1BJ, UK}
}

\begin{abstract}
I present a selection of recent lattice results in
flavourdynamics, including the status of the calculation of quark
masses and a variety of weak matrix elements relevant for the
determination of CKM matrix elements. Recent improvements in the
momentum resolution of lattice computations and progress towards
precise computations of $K\to\pi\pi$ decay amplitudes are also
reviewed.
\end{abstract}

\maketitle


\section{Introduction}
One of the main approaches to testing the Standard Model of
Particle Physics and searching for signatures of new physics is to
study a large number of physical processes to obtain information
about the unitarity triangle and to check its consistency. The
precision with which this check can be accomplished is limited by
non-perturbative QCD effects and lattice QCD provides the
opportunity to quantify these effects without model assumptions.
Of course, lattice computations themselves have a number of
sources of systematic uncertainty, and much of our current effort
is being devoted to reducing and controlling these errors. In this
talk I briefly discuss the evaluation of quark masses and weak
matrix elements using lattice simulations.

For most lattice calculations of physical quantities, the
principal source of systematic uncertainty is the \textit{chiral
extrapolation}, i.e. the extrapolation of results obtained with
unphysically large $u$ and $d$ quark masses. Ideally we would like
to perform computations with 140\,MeV pions and hence with
$m_q/m_s$ of about 1/25 (where $m_q$ ($m_s$) is the average light
quark mass (strange quark mass)). In practice values $m_q/m_s\ge$
1/2 are fairly typical, so that the MILC Collaboration's
simulation with $m_q/m_s\simeq 1/8$ is particularly
impressive~\cite{bernard} and provides a challenge to the rest of
the community to reach similarly low masses. Its configurations
have been widely used to determine physical quantities with small
quoted errors.

The MILC collaboration uses the staggered formulation of lattice
fermions and for a variety of reasons it is very important to
verify the results using other formulations. With staggered
fermions each meson comes in 16 \textit{tastes} and the unphysical
ones are removed by taking the fourth root of the fermion
determinant. Although there is no demonstration that this
procedure is wrong, there is also no proof that it correctly
yields QCD in the continuum limit~\cite{durr}. The presence of
unphysical tastes leads to many parameters to be fit in staggered
chiral perturbation theory (typically many tens of parameters) and
to date the renormalization has only been performed using
perturbation theory. It is therefore pleasing to observe that the
challenge of reaching lower masses is being taken up by groups
using other formulations of lattice fermions (see e.g.
ref.\,\cite{luscherdublin})\,.

In this talk I will discuss a selection of issues and results in
lattice flavourdynamics. I start by describing some new thoughts
on improving the momentum resolution in simulations, by varying
the boundary conditions on the quark fields. I then review the
status of lattice calculations of quark masses, $K_{\ell 3}$
decays (for which computations have only recently began) and
$B_K$. This is followed by a discussion of some of the key issues
in the computation of $K\to\pi\pi$ decays and in heavy-quark
physics.

\subsection{Improving the Momentum Resolution on the Lattice}
\label{subsec:twisted}

Numerical simulations of lattice QCD are necessarily performed on
a finite spatial volume, $V=L^3$. Providing that $V$ is
sufficiently large, we are free to choose any consistent boundary
conditions for the fields $\phi(\vec{x},t)$, and it is
conventional to use periodic boundary conditions,
$\phi(x_i+L)=\phi(x_i)$ ($i=1,2$ or 3). This implies that
components of momenta are quantized to take integer values of
$2\pi/L$. Taking a typical example of a lattice with 24 points in
each spatial direction, $L=24a$, with a lattice spacing
$a=0.1$\,fm so that $a^{-1}\simeq 2$\,GeV, we have
$2\pi/L=.52$\,GeV. The available momenta for phenomenological
studies (e.g. in the evaluation of form-factors) are therefore
very limited, with the allowed values of each component $p_i$
separated by about 1/2 GeV. The momentum resolution in such
simulations is very poor.

Bedaque~\cite{bedaque} has advocated the use of \textit{twisted}
boundary conditions for the quark fields $q(\vec{x})$ e.g.
\begin{equation}\label{eq:twisted}
q(x_i+L)=e^{i\theta_i} q(x_i)\quad\textrm{with momentum
spectrum}\quad
p_i=n_i\frac{2\pi}{L}+\frac{\theta_i}{L}\,,\end{equation} with
integer $n_i$. Modifying the boundary conditions changes the
finite-volume effects, however, for quantities which do not
involve \textit{Final State Interactions} (e.g. hadronic masses,
decay constants, form-factors) these errors remain exponentially
small also with twisted boundary conditions~\cite{giovanni}. Since
we usually neglect such errors when using periodic boundary
conditions, we can use twisted boundary conditions with the same
precision. Moreover the finite-volume errors are also
exponentially small for \textit{partially twisted boundary
conditions} in which the sea quarks satisfy periodic boundary
conditions but the valence quarks satisfy twisted boundary
conditions~\cite{giovanni,chen}. This is of significant practical
importance, implying that we do not need to generate new gluon
configurations for every choice of twisting angle $\{\theta_i$\}.

\begin{figure}
\hspace{4mm}
\begin{minipage}{.49\linewidth}
\psfrag{kappa}[c][c][1][0]{}
\psfrag{aEthetasq}[b][t][1][0]{\footnotesize$(aE(\theta))^2$}
\psfrag{pLsq}[t][c][1][0]{\footnotesize$(\vec p L)^2$}
  \epsfig{scale=.26,angle=270,file=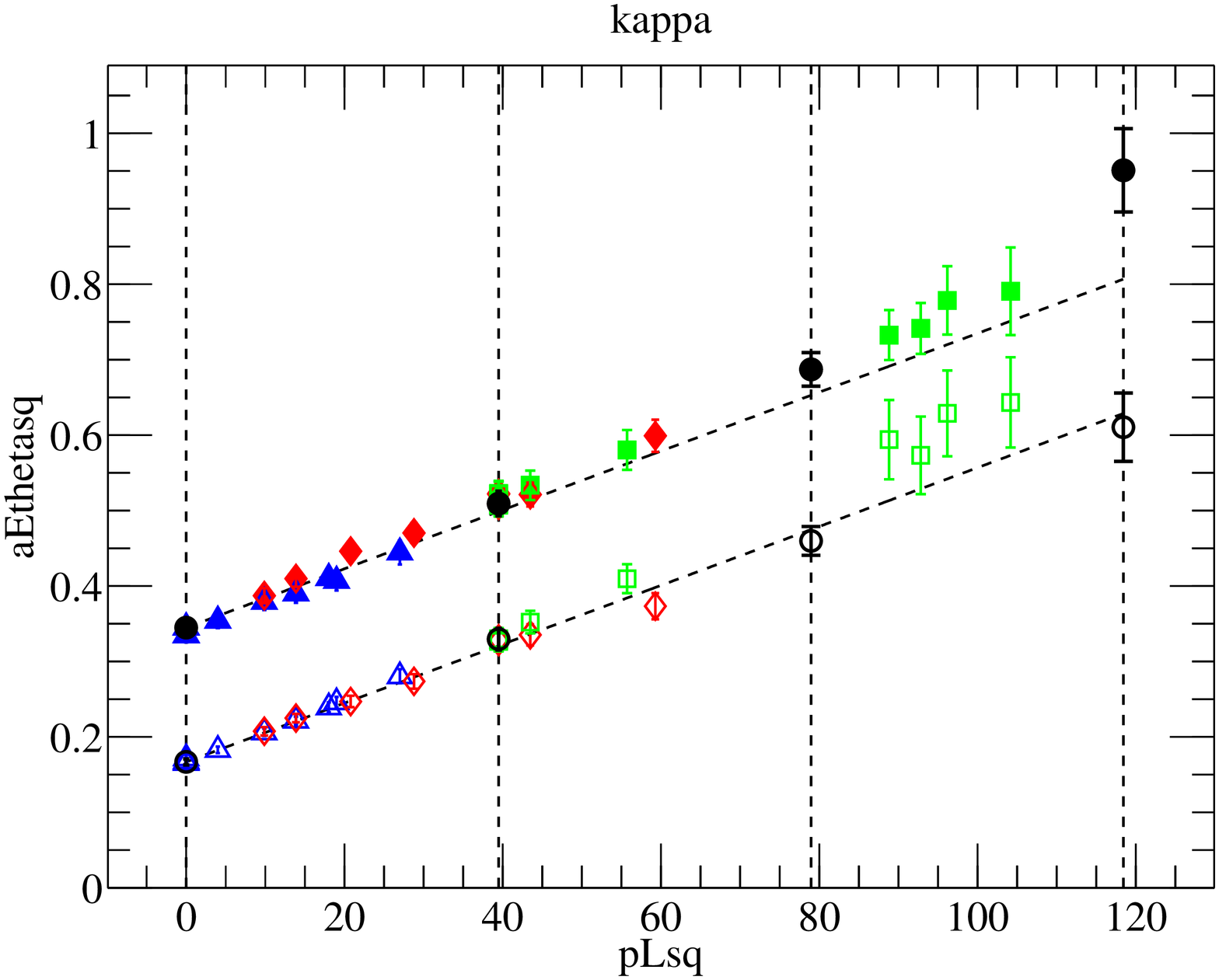}\\[2mm]
\end{minipage}
\begin{minipage}{.49\linewidth}
\psfrag{kappa}[c][c][1][0]{}
\psfrag{Fbare}[b][t][1][0]{\footnotesize$af_{\rm bare}$}
\psfrag{pLsq}[t][c][1][0]{\footnotesize$(\vec p L)^2$}
  \epsfig{scale=.26,angle=270,file=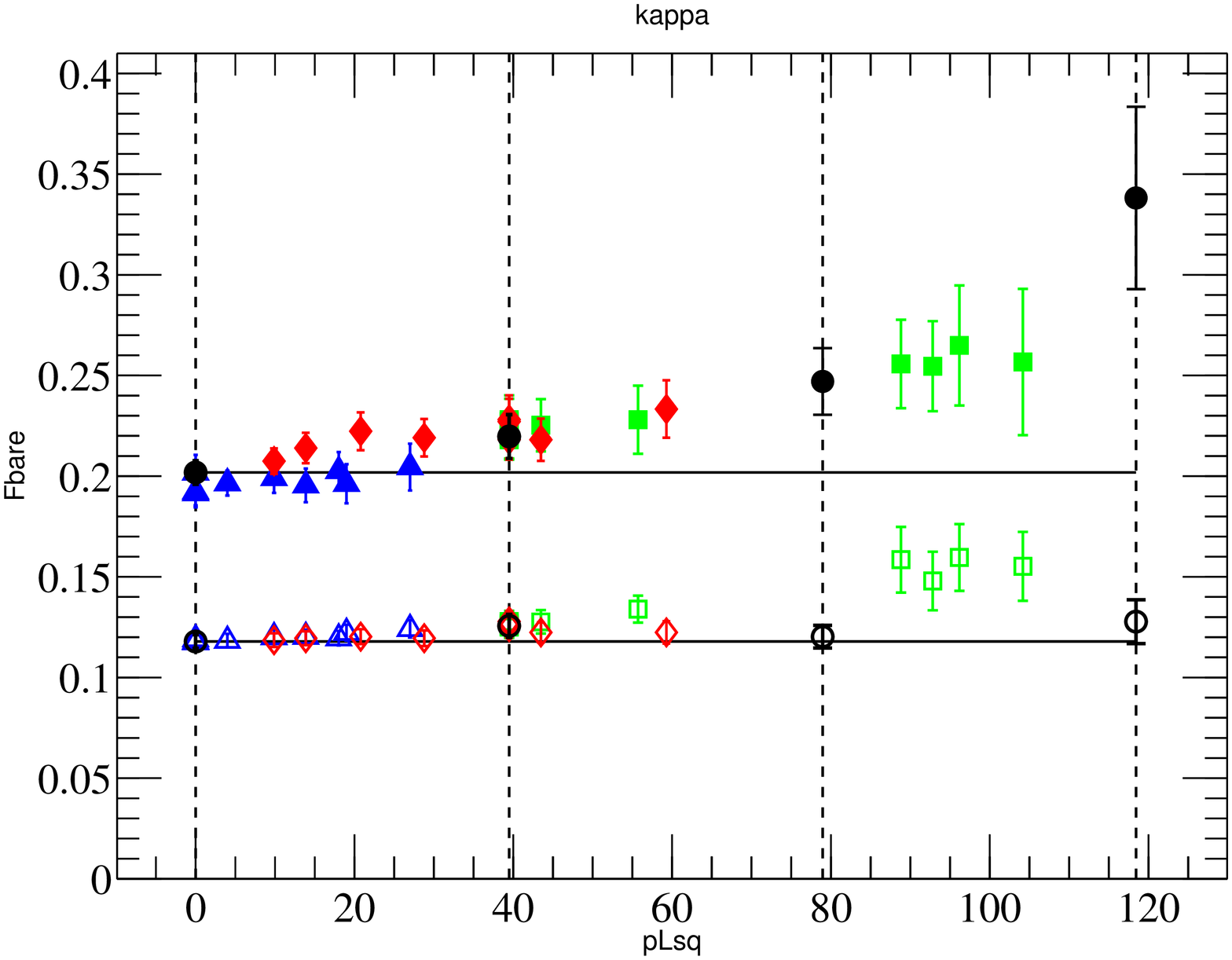}\\[2mm]
\end{minipage}
\caption{Plots of the Dispersion Relation (left) and Decay
Constants (right) as a function of the momentum $\vec{p}$ of the
mesons. In both cases the top (bottom) plot corresponds to the
$\rho$-meson ($\pi$-meson).}\label{fig:twist}
\end{figure}

The use of partially twisted boundary conditions opens up many
interesting phenomenological applications, solving the problem of
poor momentum resolution. It also appears to work numerically.
Consider for example, the plots in fig.\,\ref{fig:twist}, obtained
using an unquenched (2 flavours of sea quarks) UKQCD simulation on
a $16^3\times32$ lattice, with a spacing of about 0.1\,fm. The
plots correspond to a value for the light-quark masses for which
$m_\pi/m_\rho=0.7$ ~\cite{juettner}. The lower (upper) left-hand
plot shows the energy of the $\pi$ ($\rho$) as a function of the
momentum of the meson, and the right-hand plot shows the bare
values of the leptonic decay constants $f_\pi$ and $f_\rho$. The
$x$-axis denotes $(|\vec{p}\,|L)^2$. The results are beautifully
consistent with expectations (particularly for $pL\le 2\pi$ where
lattice artifacts are small); the predicted dispersion relation is
satisfied and the extracted decay constants are independent of the
momenta. Using periodic boundary conditions only the results at
values of $\vec{p}$ indicated by the dashed lines are accessible.
With partially twisted boundary conditions all momenta are
reachable.

\section{Quark Masses}

Quark Masses are fundamental parameters of the Standard Model, but
unlike leptons, quarks are confined inside hadrons and are not
observed as physical particles. Quark masses therefore cannot be
measured directly, but have to be obtained indirectly through
their influence on hadronic quantities and this frequently
involves non-perturbative QCD effects. Lattice simulations prove
to be very useful in the determination of quark masses;
particularly for the light quarks ($u,d$ and $s$) for which
perturbation theory is inapplicable.

In order to determine the quark masses we compute a convenient and
appropriate set of physical quantities (frequently a set of
hadronic masses) and vary the input masses until the computed
values correctly reproduce the set of physical quantities being
used for calibration. In this way we obtain the physical values of
the bare quark masses, from which by using perturbation theory, or
preferably \textit{non-perturbative renormalization}, the results
in standard continuum renormalization schemes can be determined.

My current \textit{best} estimates for the values of the quark
masses as determined from lattice simulations are presented in
table \ref{tab:quark_masses}.

\begin{table}\label{tab:quark_masses}
\begin{tabular}{|c|c|c|c|}\hline
\textbf{Flavour}&\textbf{\textit{Best} Lattice Values} &
\textbf{PDG 2004 (Lattice)}& \textbf{PDG 2004 (Non-Lattice)}\\
\hline $\hat{m}(2\,\textrm{GeV})$&($3.8\pm 0.8$)\,MeV&
($4.2\pm 1.0$)\,MeV&$(1.5<m_u(2\,\textrm{GeV})<5)$\,MeV\\
&&&$(5<m_d(2\,\textrm{GeV})<9)$\,MeV\\
$m_s(2\,\textrm{GeV})$&$(95\pm 20)\,$MeV&$(105\pm 25)\,$MeV&
80\,--\,155\,MeV\\ $\bar{m}_c$&$(1.26\pm 0.13\pm
0.20)$\,GeV&$(1.30\pm 0.03\pm 0.20)$\,GeV& 1\,--\,1.4\,GeV\\
$\bar{m}_b$&$(4.2\pm 0.1\pm 0.1)$\,GeV&$(4.26\pm 0.15\pm
0.15)$\,GeV& 4\,--\,4.5\,GeV
\\ \hline
\end{tabular}\caption{My summary of the status of lattice
determinations of quark masses (in the $\overline{\textrm{MS}}$
renormalization scheme). For $\hat{m}\equiv(m_u+m_d)/2$ and $m_s$
the results are presented at 2\,GeV and for $m_c,\,m_b$, the
results are presented at the mass itself ($\bar{m}\equiv
\bar{m}(\bar{m})$)\,. For comparison the values quoted by the PDG
in 2004, using or excluding lattice simulations, are also
presented.}
\end{table}

The relatively large error on the mass of the charm quark is a
reflection of the fact that the most detailed study to date was
performed in the quenched approximation~\cite{rolf}, whose authors
find $\bar{m}_c = 1.301(34)\,\textrm{GeV}$. I have added a
conservative 15\% error as an estimate of quenching effects.
Current and future calculations will be dominated by unquenched
simulations so that the error will decrease very significantly.
Indeed a very recent unquenched calculation finds
$\bar{m}_c=1.22(9)$\,GeV~\cite{nobes}.

The relative error on $m_b$ is small because what is actually
calculated is $m_B-m_b$. The calculations are performed in the
Heavy Quark Effective Theory and the major source of systematic
error is the subtraction of $O(1/(a\Lambda_{\textrm{QCD}}))$
terms. Using stochastic perturbation theory, Di Renzo and Scorzato
have performed this calculation to 3-loop order~\cite{direnzo}.
The second error on $\bar{m}_b$ in table~\ref{tab:quark_masses} is
my conservative estimate of the fact that the simulations have
been performed with two flavours of sea quarks.

\section{Selected Topics in Kaon Physics}

\subsection{\boldmath{$K_{\ell 3}$} Decays}

A new area of investigation for lattice simulations is the
evaluation of non-perturbative QCD effects in
$K\to\pi\ell\nu_\ell$ decays, from which the CKM matrix element
$V_{us}$ can be determined. The QCD contribution to the amplitude
is contained in two invariant form-factors $f^0(q^2)$ and
$f^+(q^2)$ defined by
\[
\langle\,\pi(p_\pi)\,|\bar{s}\gamma_\mu u\,|\,K(p_K)\,\rangle=
f^0(q^2)\,\frac{M_K^2-M_\pi^2}{q^2}q_\mu +
f^+(q^2)\,\left[(p_\pi+p_K)_\mu-\frac{M_K^2-M_\pi^2}{q^2}q_\mu
\right]\,,
\]
where $q=p_K-p_\pi$. (Parity Invariance implies that only the
vector current from the $V-A$ charged current contributes to the
decay.) A useful reference value for $f^+(0)$ comes from the
20-year old prediction of Leutwyler and Roos,
$f^+(0)=1+f_2+f_4+\cdots=0.961(8)$ where
$f_n=O(M^2_{K,\pi,\eta})$. $f_2=-0.023$ is well determined,
whereas the higher order terms in the chiral expansion require
model assumptions.

To be useful in extracting $V_{us}$ from experimental measurements
we need to be able to evaluate $f^0(0)=f^+(0)$ to better than
about 1\% precision. This would seem to be impossible until one
notes that it is possible to compute $1-f^+(0)$, so that an error
of 1\% on $f^+(0)$ is actually an error of O(25\%) on $1-f^+(0)$.
The calculation follows a similar strategy to that proposed in
ref.~\cite{hashimoto} for the form-factors of $B\to D$
semileptonic decays (which in the heavy quark limit are also close
to 1), starting with a computation of double ratios such as
\begin{equation}
\frac{\langle\pi|\bar{s}\gamma_0 l|K\rangle \langle
K|\bar{l}\gamma_0 s|\pi\rangle}{\langle\pi|\bar{l}\gamma_0
l|\pi\rangle \langle K|\bar{s}\gamma_0
s|K\rangle}=\left[f^0(q^2_{\textrm{max}})\right]^2\,\frac{(m_K+m_\pi)^2}{4m_Km_\pi}\,,
\end{equation}
where all the mesons are at rest and
$q^2_{\textrm{max}}=(M_K-M_\pi)^2$.

Following a quenched calculation by the SPQR collaboration last
year~\cite{spqrkl3}, in which the strategy for determining the
form-factors was presented, there have been 3 very recent
unquenched (albeit largely preliminary) results:
\begin{center}\vspace{-0.2in}\begin{tabular}{rl}\\
\textrm{RBC~\cite{rbckl3}} & $f^+(0)$\,=\,0.955
(12)\\
\textrm{JLQCD~\cite{jlqcdkl3}} & $f^+(0)$\,=\,0.952\,(6)\\
\textrm{FNAL/MILC/HPQCD~\cite{milckl3}} &
$f^+(0)$\,=\,0.962\,(6)\,(9)
\end{tabular}\end{center}
in good agreement with the result of Leutwyler and
Roos~\cite{leutwyler}.

\subsection{\boldmath{$B_K$}}

$B_K$, the parameter which contains the non-perturbative QCD
effects in $K^0-\bar{K}^0$ mixing, has been computed in lattice
simulations by many groups. It is defined by
\begin{equation}
\langle\bar{K}^0\,|\,(\bar{s}\gamma^\mu(1-\gamma^5)d)\,
(\bar{s}\gamma_\mu(1-\gamma^5)d)|K^0\rangle=\frac83
\,M_{K}^2\,f_{K}^2\,B_{K}\,.\end{equation} $B_K$ depends on the
renormalization scheme and scale and is conventionally given in
the NDR, $\overline{\textrm{MS}}$ scheme at $\mu=2\,$GeV or as the
RGI parameter $\hat{B}_K$ (the relation between the two is
$\hat{B}_K\simeq 1.4\ B_K^{\overline{\textrm{MS}}}
(2\,\textrm{GeV})$).

\begin{figure}\label{fig:bk}
\includegraphics
[width=0.55\hsize] {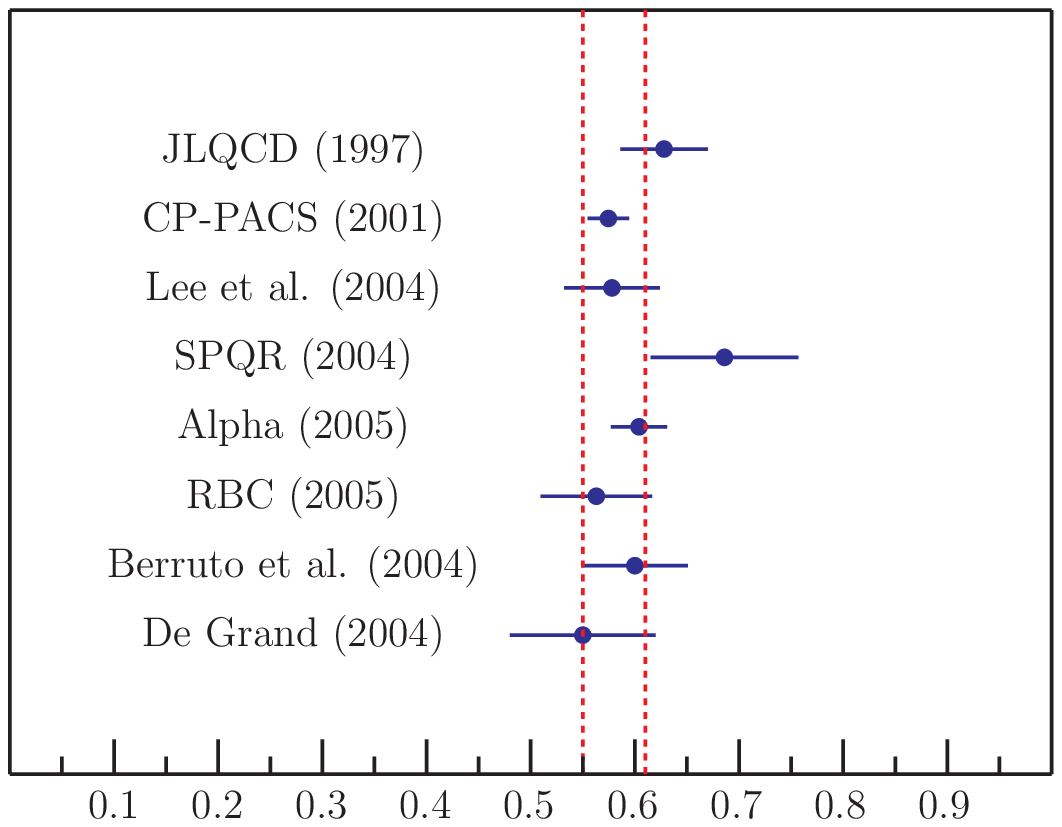} \caption{A compilation of recent
quenched results for $B^{\overline{MS}}_K(2\,\textrm{GeV})$.}
\end{figure}

A compilation of recent results for $B_K$ obtained in the quenched
approximation is presented in fig.\,\ref{fig:bk}. From such
results recent reviewers have summarised the status of quenched
calculations as:
\begin{equation}
B^{\overline{MS}}_K(2\,\textrm{GeV})=0.58(4)~\textrm{\cite{hashimotoichep}}\quad
\textrm{and}\quad B^{\overline{MS}}_K(2\,\textrm{GeV})=0.58(3)
~\textrm{\cite{dawsondublin}}\,.
\end{equation}
The dashed lines in fig.\,\ref{fig:bk} correspond to $B_K=.58(3)$,
which I am happy to take as the current best estimate.

The challenge now is to obtain reliable unquenched results; such
computations are underway by several groups but so far the results
are very preliminary. We will have to wait a year or two for
precise results, but I mention in passing C.Dawson's
guesstimate~\cite{dawsondublin} (stressing that it is only a
guesstimate), based on a comparison of quenched and unquenched
results at similar masses and lattice spacings, of
$B^{\overline{MS}}_K(2\,\textrm{GeV})=0.58(3)(6)$.

\subsection{\boldmath{$K\to\pi\pi$} Decays}

A quantitative understanding of non-perturbative effects in
$K\to\pi\pi$ decays will be an important future milestone for
lattice QCD. Two particularly interesting challenges are:\\
i) an understanding of the empirical $\Delta I=1/2$ rule, which
states that the amplitude for decays in which the two-pion final
state has isospin I=0 is larger by a factor of about 22 than
that in which the final state has $I=2$;\\
ii) a calculation of $\varepsilon^\prime/\varepsilon$, whose
experimental measurement with a non-zero value, $(17.2\pm
1.8)\times 10^{-4}$, was the first observation of direct
CP-violation.\\
The two challenges require the computation of the matrix elements
of the $\Delta S=1$ operators which appear in the effective Weak
Hamiltonian.

About 4 years ago, two collaborations published some very
interesting quenched results for these quantities:
\begin{center}
\begin{tabular}{|c|c|c|}\hline Collaboration(s)&{\textrm Re}
$A_0$/{\textrm Re} $A_2$ & $\varepsilon^\prime/\varepsilon$\\
\hline
RBC~\cite{rbckaon} & $25.3\pm 1.8$ &$-(4.0\pm 2.3)\times 10^{-4}$\\
CP-PACS\cite{cppacskaon}
& 9\,--\,12&(-7\,--\,-2)$\times 10^{-4}$\\
Experiments & 22.2 & $(17.2\pm 1.8)\times 10^{-4}$\\
\hline
\end{tabular}\end{center}
Both collaborations obtain a considerable octet enhancement
(significantly driven however, by the chiral extrapolation) and
$\varepsilon^\prime/\varepsilon$ with the wrong sign. A
particularly impressive feature of these calculations was that the
collaborations were able to perform the subtraction of the
unphysical terms which diverge as powers of the ultra-violet
cut-off ($a^{-1}$, where $a$ is the lattice spacing). The results
are very interesting and will provide valuable benchmarks for
future calculations, however the limitations of the calculations
should be noted, in particular the use of chiral perturbation
theory ($\chi$PT) only at lowest order. This has the practical
advantage that $K\to\pi\pi$ matrix elements do not have to be
evaluated directly, it is sufficient at lowest order to study the
mass dependence of the matrix elements $\langle\,M\,|\,{\cal
O}_i\,|\,M\,\rangle$ and $\langle\,0\,|\,{\cal
O}_i\,|\,M\,\rangle$, where $M$ is a pseudoscalar meson and the
${\cal O}_i$ are the $\Delta S=1$ operators appearing in the
effective Hamiltonian, to determine the low-energy constants and
hence the amplitudes. It is not very easy to estimate the errors
due to this approximation, but they should be at least of
$O(m_K^2/\Lambda_{\textrm{{\small QCD}}}^2)$. Since for
$\varepsilon^\prime/\varepsilon$ the dominant contributions appear
to be from the QCD and electroweak penguin operators ${\cal O}_6$
and ${\cal O}_8$, which are comparable in magnitude but come with
opposite signs, it is not totally surprising that the prediction
for $\varepsilon^\prime/\varepsilon$ at lowest order in $\chi$PT
has the wrong sign. It should also be noted that in the
simulations described in ref.~\cite{rbckaon,cppacskaon} the light
quarks masses were large (the pions were heavier than about
400\,MeV) and so one can question the validity of $\chi$PT in the
range of masses used (about 400-800\,MeV).

To improve the precision, apart from performing unquenched
simulations and reducing the masses of the light quarks, one needs
to go beyond lowest order $\chi$PT (for example by going to
NLO~\cite{lin,laihosoni}) and, in general, this requires the
evaluation of $K\to\pi\pi$ matrix elements and not just $M\to M$
ones. The treatment of two-hadron states in lattice computations
has a new set of theoretical issues, most notably the fact that
the finite-volume effects decrease only as powers of the volume
and not exponentially. Starting with the pioneering work of
L\"uscher~\cite{luscher}, the theory of finite-volume effects for
two-hadron states in the elastic regime is now fully understood,
both in the centre-of-mass and moving
frames,~\cite{luscher}\,--\,\cite{cky} and I will now briefly
discuss this.

Consider the two-hadron correlation function represented by the
diagram
\begin{center}
\begin{picture}(280,30)(-10,-10)
\Oval(30,0)(10,20)(0)\Oval(70,0)(10,20)(0)
\Oval(110,0)(10,20)(0)\Oval(150,0)(10,20)(0)
\Oval(190,0)(10,20)(0)\Oval(230,0)(10,20)(0)%
\Line(-10,10)(10,0)\Line(-10,-10)(10,0)
\Line(250,0)(270,10)\Line(250,0)(270,-10)
\GCirc(10,0){5}{0.75}\GCirc(50,0){5}{0.75} \GCirc(90,0){5}{0.75}
\GCirc(130,0){5}{0.75}
\GCirc(170,0){5}{0.75}\GCirc(210,0){5}{0.75}
\GCirc(250,0){5}{0.75}%
\ArrowLine(149.5,10)(150.5,10)\Text(150,16)[b]{$p$}
\Text(-10,0)[r]{$E$}
\end{picture}
\end{center}
where the shaded circles represent two-particle irreducible
contributions in the $s$-channel. For simplicity let us take the
two-hadron system to be in the centre-of mass frame and assume
that only the $s$-wave phase-shift is significant (the discussion
can be extended to include higher partial waves). Consider the
loop integration/summation over $p$ (see the figure). Performing
the $p_0$ integration by contours, we obtain a summation over the
spatial momenta of the form:
\begin{equation}\label{eq:fv}
\frac{1}{L^3}\sum_{\vec{p}}\frac{f(p^2)}{p^2-k^2}
\end{equation}
where the relative momentum $k$ is related to the energy by
$E^2=4(m^2+k^2)$, the function $f(p^2)$ is non-singular and (for
periodic boundary conditions) the summation is over momenta
$\vec{p}=(2\pi/L)\,\vec{n}$ where $\vec{n}$ is a vector of
integers. In infinite volume the summation in eq.\,(\ref{eq:fv})
is replaced by an integral and it is the difference between the
summation and integration which gives the finite-volume
corrections. The relation between finite-volume sums and
infinite-volume integrals is the \textit{Poisson Summation
Formula}, which (in 1-dimension) is:
\begin{equation}\label{eq:poisson}
\frac{1}{L}\sum_{p}g(p)=\sum_{l=-\infty}^\infty\int\frac{dp}
{2\pi}e^{ilLp}g(p)\,.
\end{equation}
If the function $g(p)$ is non-singular, the oscillating factors on
the right-hand side ensures that only the term with $l=0$
contributes, up to terms which vanish exponentially with $L$. The
summand in eq.\,(\ref{eq:fv}) on the other hand is singular (there
is a pole at $p^2=k^2$) and this is the reason why the
finite-volume corrections only decrease as powers of $L$. The
detailed derivation of the formulae for the finite-volume
corrections can be found in refs.\,~\cite{luscher}\,--\,\cite{cky}
and is beyond the scope of this talk. The results hold not only
for $K\to\pi\pi$ decays, but also for $\pi$\,-\,nucleon and
nucleon-nucleon systems.

For decays in which the two-pions have isospin 2, we now have all
the necessary techniques to calculate the matrix elements with
good precision and such computations are underway. For decays into
two-pion states with isospin 0 there are also no barriers in
principle. However, in this case, purely gluonic intermediate
states contribute and we need to learn how to calculate the
corresponding disconnected diagrams with sufficient precision. In
addition the subtraction of power-like ultraviolet divergences
requires large datasets (as demonstrated in refs.\,\cite{rbckaon,
cppacskaon} in quenched QCD). For these reasons it will take a
longer time for some of the $\Delta I=1/2$ matrix elements to be
computed than $\Delta I=3/2$ ones.

\section{Heavy Quark Physics}
Lattice simulations are playing an important role in the
determination of physical quantities in heavy quark physics
including decay constants ($f_B,f_{B_s},f_D,f_{D_s}$), the
$B$-parameters of $B-\bar{B}$ mixing (from which the CKM matrix
elements $V_{td}$ and $V_{ts}$ can be determined), form-factors of
semileptonic decays (which give $V_{cb}$ and $V_{ub}$), the
$g_{BB^\ast\pi}$ coupling constant of heavy-meson chiral
perturbation theory and the lifetimes of beauty hadrons.

The typical lattice spacing in current simulations $a\simeq
0.1$\,fm is larger than the Compton wavelength of the $b$-quark
and comparable to that of the $c$-quark. The simulations are
therefore generally performed using effective theories, such as
the Heavy Quark Effective Theory or Non-Relativistic QCD. Another
interesting approach was proposed by the Fermilab
group~\cite{fnal}, in which the action is \textit{improved} to the
extent that, in principle at least, artefacts of $O((m_Qa)^n)$ are
eliminated for all $n$, where $m_Q$ is the mass of the heavy quark
$Q$. Determining the coefficients of the operators in these
actions requires matching with QCD, and this matching is almost
always performed using perturbation theory (most often at one-loop
order). This is a significant source of uncertainty and provides
the motivation for attempts to develop non-perturbative matching
techniques.

I only have time here to consider very briefly a single topic,
semileptonic $B$-decays. For $B\to\pi$ decays, the pion's momentum
has to be small in order to avoid large lattice artefacts, so that
$q^2=(p_B-p_\pi)^2$ is large ($q^2>15$\,GeV$^2$ or so). There
continues to be a considerable effort in extrapolating these
results over the whole $q^2$ range. Recently, as experimental
results begin to be presented in $q^2$ bins, it has become
possible to combine the lattice results at large $q^2$ with the
binned experimental results and theoretical constraints to obtain
$V_{ub}$ with good precision~\cite{becher}.

As an example I present a recent result, obtained using the MILC
gauge field configurations with staggered light quarks
and the Fermilab action for the $b$-quark~\cite{fnalbtopi}\\
\begin{equation}
|V_{ub}|=3.48(29)(38)(47)\times 10^{-3}\,.
\end{equation}
I mention that other semileptonic decays of heavy mesons are also
being studied, including $B\to D^{(\ast)}$ decays (a recent result
is $|V_{cb}|=3.9(1)(3)\times 10^{-2}$~\cite{milckl3}) and
$D\to\pi,K$ decays.

\section{Summary and Conclusions}
Lattice QCD simulations, in partnership with experiments and
theory, play a central r\^ole in the determination of the
fundamental parameters of the Standard Model (e.g. quark masses,
CKM matrix elements) and in searches for signatures of new physics
and ultimately perhaps will help to unravel its structure. With
the advent of unquenched simulations, a major source of
uncontrolled systematic uncertainty has been eliminated and the
main aim now is to control the chiral extrapolation and reduce
other systematic uncertainties. We continue to extend the range of
applicability of lattice simulations to more processes and
physical quantities. In this talk I have only been able to give a
small selection of recent results and developments; a more
complete set can be found on the web-site of the 2005
international symposium on lattice field theory~\cite{lat05}.




\bibliographystyle{aipprocl} 
\vspace{-0.1in}


\begin{thebibliography}{99}
\bibitem{bernard}
C.~Bernard {\it et al.}  [MILC Collaboration],
PoS {\bf LAT2005} (2005) 025 [arXiv:hep-lat/0509137] and
references therein.
\bibitem{durr}
S.~Durr,
PoS {\bf LAT2005} (2005) 021 [arXiv:hep-lat/0509026].
\bibitem{luscherdublin}
M.~Luscher,
PoS {\bf LAT2005} (2005) 002 [arXiv:hep-lat/0509152]; L.~Del
Debbio, L.~Giusti, M.~Luscher, R.~Petronzio and N.~Tantalo,
arXiv:hep-lat/0512021.
\bibitem{bedaque}
P.~F.~Bedaque,
\emph{Phys.\ Lett.\ B} {\bf 593} 82 (2004)
[arXiv:nucl-th/0402051].%
\bibitem{giovanni}
C.~T.~Sachrajda and G.~Villadoro,
\emph{Phys.\ Lett.\ B} {\bf 609} 73 (2005)
[arXiv:hep-lat/0411033].
\bibitem{chen}
P.~F.~Bedaque and J.~W.~Chen,
\emph{Phys.\ Lett.\ B} {\bf 616} 208 (2005)
[arXiv:hep-lat/0412023].
\bibitem{juettner}
J.~M.~Flynn, A.~Juttner and C.~T.~Sachrajda  [UKQCD
Collaboration],
\emph{Phys.\ Lett.\ B} {\bf 632} 313 (2006)
[arXiv:hep-lat/0506016]; J.~Flynn, A.~Juttner, C.~Sachrajda and
G.~Villadoro,
PoS {\bf LAT2005} (2005) 352 [arXiv:hep-lat/0509093].
\bibitem{rolf}
J.~Rolf and S.~Sint  [ALPHA Collaboration],
\emph{JHEP} {\bf 0212} 007 (2002) [arXiv:hep-ph/0209255].
\bibitem{nobes}
M.~Nobes and H.~Trottier,
PoS {\bf LAT2005} (2005) 209 [arXiv:hep-lat/0509128].
\bibitem{direnzo}
F.~Di Renzo and L.~Scorzato,
\emph{Nucl.\ Phys.\ Proc.\ Suppl.\ }{\bf 140} 473 (2005)
[arXiv:hep-lat/0409151].
\bibitem{hashimoto}
S.~Hashimoto, A.~X.~El-Khadra, A.~S.~Kronfeld, P.~B.~Mackenzie,
S.~M.~Ryan and J.~N.~Simone,
\emph{Phys.\ Rev.\ D} {\bf 61} 014502 (2000)
[arXiv:hep-ph/9906376].
\bibitem{leutwyler}
H.~Leutwyler and M.~Roos,
\emph{Z.\ Phys.\ C} {\bf 25} 91 (1984).
\bibitem{spqrkl3}
D.~Becirevic {\it et al.},
\emph{Nucl.\ Phys.\ B} {\bf 705} 339 (2005)
[arXiv:hep-ph/0403217].
\bibitem{rbckl3}
C.~Dawson \textit{et.al.},
PoS {\bf LAT2005} (2005) 337 [arXiv:hep-lat/0510018].
\bibitem{jlqcdkl3}
N.~Tsutsui {\it et al.}  [JLQCD Collaboration],
PoS {\bf LAT2005} (2005) 357 [arXiv:hep-lat/0510068].
\bibitem{milckl3}
M.~Okamoto  [Fermilab Lattice Collaboration],
arXiv:hep-lat/0412044. 
\bibitem{hashimotoichep}
S.~Hashimoto,
\emph{Int.\ J.\ Mod.\ Phys.\ A} {\bf 20} 5133 (2005)
[arXiv:hep-ph/0411126].
\bibitem{dawsondublin}
C.Dawson,
\texttt{http://www.maths.tcd.ie/lat05/plenary-talks/Dawson.pdf}
\bibitem{rbckaon}
T.~Blum {\it et al.}  [RBC Collaboration],
\emph{Phys.\ Rev.\ D} {\bf 68} 114506 (2003)
[arXiv:hep-lat/0110075].
\bibitem{cppacskaon}
J.~I.~Noaki {\it et al.}  [CP-PACS Collaboration],
\emph{Phys.\ Rev.\ D} {\bf 68} 014501 (2003)
[arXiv:hep-lat/0108013].
\bibitem{lin}
C.~J.~D.~Lin \textit{et.al.},
\emph{Nucl.\ Phys.\ B} {\bf 650} 301 (2003)
[arXiv:hep-lat/0208007]; P.~Boucaud {\it et al.}  [The SPQ(CD)R
Collaboration],
\emph{Nucl.\ Phys.\ Proc.\ Suppl.\ }{\bf 106} 329 (2002)
[arXiv:hep-lat/0110206].
\bibitem{laihosoni}
J.~Laiho and A.~Soni,
\emph{Phys.\ Rev.\ D }{\bf 65} 114020 (2002)
[arXiv:hep-ph/0203106];
\emph{Phys.\ Rev.\ D} {\bf 71} 014021 (2005)
[arXiv:hep-lat/0306035].
\bibitem{luscher}
M.~Luscher,
\emph{Commun.\ Math.\ Phys.}\  {\bf 104} 177 (1986);
\emph{Commun.\ Math.\ Phys.}\  {\bf 105} 153 (1986);
\emph{Nucl.\ Phys.\ B} {\bf 354} 531 (1991);
\emph{Nucl.\ Phys.\ B} {\bf 364} 237 (1991).
\bibitem{rummukainen}
K.~Rummukainen and S.~A.~Gottlieb,
\emph{Nucl.\ Phys.\ B} {\bf 450} 397 (1995)
[arXiv:hep-lat/9503028].
\bibitem{ll}
L.~Lellouch and M.~Luscher,
\emph{Commun.\ Math.\ Phys.}\ {\bf 219} 31 (2001)
[arXiv:hep-lat/0003023].
\bibitem{lmst}
C.~J.~D.~Lin, G.~Martinelli, C.~T.~Sachrajda and M.~Testa,
\emph{Nucl.\ Phys.\ B} {\bf 619} 467 (2001)
[arXiv:hep-lat/0104006].
\bibitem{kss}
C.~h.~Kim, C.~T.~Sachrajda and S.~R.~Sharpe,
\emph{Nucl.\ Phys.\ B} {\bf 727} (2005) 218
[arXiv:hep-lat/0507006]; PoS {\bf LAT2005} (2005) 359
[arXiv:hep-lat/0510022].
\bibitem{cky}
N.~H.~Christ, C.~Kim and T.~Yamazaki,
\emph{Phys.\ Rev.\ D} {\bf 72} 114506 (2005)
[arXiv:hep-lat/0507009].
\bibitem{fnal}
A.~X.~El-Khadra, A.~S.~Kronfeld and P.~B.~Mackenzie,
\emph{Phys.\ Rev.\ D} {\bf 55} 3933 (1997)
[arXiv:hep-lat/9604004].
\bibitem{becher}
T.~Becher and R.~J.~Hill,
arXiv:hep-ph/0509090.
\bibitem{fnalbtopi}
M.~Okamoto,
arXiv:hep-ph/0505190.
\bibitem{lat05}
\texttt{www.maths.tcd.ie/lat05}
\end{thebibliography}

\end{document}